\documentclass[%
 reprint,
 amsmath,amssymb,
 aps,
]{revtex4-2}

\usepackage{graphicx}
\usepackage{float}
\usepackage{dcolumn}
\usepackage{bm}
\usepackage{diagbox}
\usepackage{array}
\usepackage{appendix}
\usepackage{gensymb}
\usepackage[colorlinks,
            linkcolor=blue,       
            anchorcolor=blue,  
            citecolor=blue,        
            ]{hyperref}
\usepackage{multirow}
\newcommand\KeyWord[1]{\textbf{Keywords}: #1}

\begin{document}

\preprint{APS/123-QED}

\title{Dual-polarization huge photonic spin Hall shift and deep-subwavelength sensing based on topological singularities in one-dimensional photonic crystals}%

\author{Yufu Liu$^{1,3}$}

\author{Xianjun Wang$^{1,3}$}%

\author{Yunlin Li$^{1,3}$}%

\author{Haoran Zhang$^{1,3}$}%

\author{Langlang Xiong$^{2,3}$}%

\author{Xingchao Qi$^{2,3}$}%

\author{Zhen Lai$^{1,3}$}%

\author{Xuezhi Wang$^{1,3}$}%

\author{Xunya Jiang$^{1,2,3}$}%
 \email{jiangxunya@fudan.edu.cn}

\affiliation{
 $^{1}$Department of Illuminating Engineering and Light Sources, School of Information Science and Engineering, Fudan University, Shanghai, 200433, China
}
 \affiliation{
 $^{2}$Institute of Future Lighting, Academy for engineering and technology, Fudan University, Shanghai, 200433, China
}
\affiliation{
 $^{3}$Engineering Research Center of Advanced Lighting Technology, Fudan University, Ministry of Education, Shanghai, 200433, China
}

\date{\today}

\begin{abstract}
 Although several efforts have been taken to enhance the photonic spin Hall shift in deep-subwavelength region, according to effective medium theory, the fundamental confliction between near-zero reflection coefficient and near-zero incident angle still hinders the further application. Here, we reveal a fundamental breakdown of effective medium theory due to the existing of topological singularity in deep-subwavelength region in one-dimensional photonic crystals. We find that near the topological singularity, huge photonic spin Hall shift can be achieved for s-polarization and p-polarization. At the topological singularity, the reflected filed is split as dipole-like distribution with zero photonic spin Hall shift for both-polarizations, which is resulted from the interfere of the spin-maintained normal light and spin-flipped abnormal light. Based on the theoretical research, dual-polarizations thickness and dielectric constant sensing devices can be designed in deep-subwavelength region. Further more, by applying more complicated layered structure, multi-channels dual-polarizations detection and broadband dual-polarizations huge spin Hall shift platform can be designed. This work paves the way to exploring the topological properties and polarization control of photonic crystals and provides a prospective method for the design of multi-channels sensitive detection spin optical devices.
\end{abstract}

\maketitle

\KeyWord{topological singularity, photonic spin Hall shift, effective medium theory}

\section{\label{sec:introduce}INTRODUCTION}
The spin-orbital interaction of light is of great importance in modern photonics, as it generates mutual conversions between spin angular momentum and orbital angular momentum \cite{o2014spin, bliokh2015spin, bliokh2015transverse, pan2016strong, ni2021multidimensional}. One types of spin-orbital interaction, a conversion from spin angular momentum to extrinsic orbital angular momentum, resulting in photonic spin-Hall effect \cite{bliokh2006conservation, bliokh2013goos, ling2017recent, xiang2017enhanced, kim2019observation}. The photonic spin-Hall effect has been widely investigated both theoretically \cite{onoda2004hall, bliokh2008geometrodynamics, rong2020photonic} and experimentally \cite{hosten2008observation, qin2009measurement, dai2020direct} in recent years, in which a spatially confined light beam (such as Gaussian beam) splits into two spin components (i.e., left-handed and right-handed) in the direction perpendicular to the plane of incidence when reflected or refracted from an interface, which is so-called as the photonic spin-Hall shift (PSHS). Take p-polarization incidence as an example, the transverse PSHS for reflected light can be approximately written as \cite{luo2011enhanced, wang2013spin}:

\begin{equation}\label{eq(1)}
  \delta = -\frac{\lambda}{2\pi}[1+\frac{|r_s|}{|r_p|}\cos(\phi_s-\phi_p)]\cot\theta_i
\end{equation}
where, $\phi_{s/p}$ is the reflected phase of s/p-polarization and $\theta_i$ is the incident angle. Eq.\eqref{eq(1)} remains applicable to s-polarization as long as the roles of $s$ and $p$ are interchanged. Eq.\eqref{eq(1)} indicates that, to get an PSHS as large as possible, the following two features should be satisfied: (i) $|r_p|\rightarrow 0$; (ii) $\theta_i$ is near-normally incident. 

Generally, the tiny spin-dependent displacements in photonic spin-Hall effect are too small to detect directly. Several efforts have been taken to satisfy these conditions to enhance the photonic spin-Hall displacements and efficiency in PSHE \cite{ling2012enhanced, takayama2018photonic, takayama2018enhanced, chen2021high, kim2021spin, kim2022reaching}. Unfortunately, these two requirements in are conflicting with each other in deep-subwavelength region, where the wavelength is much larger than the scale of structure. In the deep-subwavelength region, the effective medium theory (EMT) is widely regarded as an approximation method used to replacing complex inhomogeneous structure as a simple homogeneous material that provides the same response \cite{lalanne1996effective, koschny2004effective, lemoult2013wave, choy2015effective, zhang2015effective}. According to EMT, condition (i) $|r_p|\rightarrow0$ requires that the incident angle $\theta_i$ is large enough, which contradicts with condition (ii). This confliction is also applied for transmitted light. As a result, it is fundamentally difficult to observe large PSHS in deep-subwavelength region experimentally. 

A few years ago, in one-dimensional (1D) photonic crystals (PhCs) with inversion symmetry, the concept and evolution of topological singularities, which can be directly positioned by the zero-scattering condition, has been extensively researched \cite{xiao2014surface, li2019two, li2019singularity, xiong2021resonance, liu2023evolution, liunearly}. The topological singularity is regarded as a topological charge, where there is a phase jump near the singularity \cite{xiao2014surface}, so that it indicates nontrivial topology of a band. In inversion symmetry gauge, the topological singularity can be directly positioned by the zero-scattering condition, which means zero-reflection of period structure. Very recently, it was found that, the topological singularity can be achieved in deep-subwavelength region when normally incident, which breaks the EMT \cite{xiong2021resonance, liu2023evolution}. This counterintuitive phenomenon means that zero reflection (i.e., $|r_{s,p}|\rightarrow0$) with near-zero incident angle $\theta_i$ can be achieved in the deep-subwavelength region, which will provide a perfect platform to research photonic spin-Hall effect.

In this work, we reveal a fundamental breakdown of EMT due to the existing of topological singularity in deep-subwavelength region in 1D PhCs with inversion symmetry. We find that near the topological singularity in the deep-subwavelength region, huge PSHS can be achieved for s-polarization and p-polarization, which breaks the EMT. Besides, at the topological singularity, the reflected filed is split as dipole-like distribution with zero PSHS for both polarizations, which results from the interfere of the spin-maintained normal light and spin-flipped abnormal light. Based on this exotic property, dual-polarizations thickness and dielectric constant sensing device can be designed, which can be applied into a wide range of fields, such as the sub-nanometer detection, non-linear power detection and environmental sensors. Finally, by applying more complicated layers, multi-channels dual-polarizations detections and dual-polarizations broadband huge-PSHS platform can be designed. Theoretically, this work builds the bridge between topology study of 1D PhCs and light-matter interaction in deep-subwavelength region. In practice, our model provides an unprecedented platform to research photonic spin-Hall effect, which can been applied in many applications, such as optical metrology, sensing and super-resolution imaging.

\section{\label{sec:2} Deep-subwavelength Dual-polarization photonic spin Hall shift}
\begin{figure*}[htb]
    \centering
    \includegraphics[width=1\linewidth]{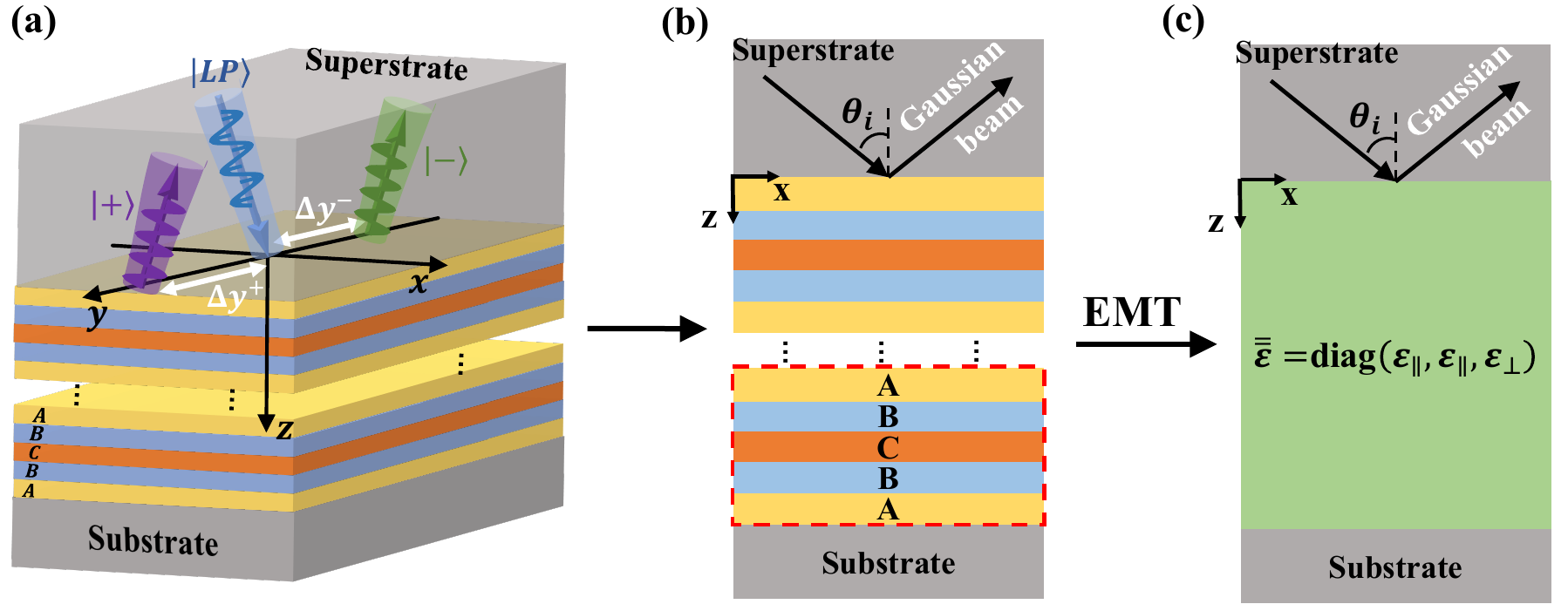}
    \caption{(a) The schematic of a p-polarized Gaussian beam reflects from 1D PhCs. The reflected beams are composed of left-handed ($|+\rangle$) and right-handed ($|-\rangle$) circular polarization with PSHS $\Delta y^+$  and $\Delta y^-$. (b) The schematic of PhCs in x-z view with periodic all-dielectric multi-layer constituents composed of three dielectrics A, B and C. The lattice constant of PhCs is $a$ and cell number is fixed as $N = 8$. (c) The schematic of homogenized effective medium with relative permittivity $\bar{\bar{\varepsilon}} = [\varepsilon_\parallel, \varepsilon_\parallel, \varepsilon_\perp]$. Below and after, we set wavelength $\lambda >> a$ and $\lambda > Na/2$. The waist radius of Gaussian beam is set as: $w_0 = 100\lambda$.}
    \label{Fig(1)}
\end{figure*}

Fig.(\ref{Fig(1)}a) shows the principal illustration of our model with a linearly polarized (LP) Gaussian beam reflects from an 1D finite PhCs with cell number $N$. The reflected beams are composed of left-handed circular polarized (CP) Gaussian beam (marked by $|-\rangle$) with PSHS $\Delta y^-$ and right-handed CP Gaussian beam (marked by $|+\rangle$) with PSHS $\Delta y^+$. The PhCs is supported by a dielectric substrate with a relative permittivity of $\varepsilon_{sub}$ and covered by a dielectric superstrate with a relative permittivity of $\varepsilon_{super}$. Below and after, we set $\varepsilon_{sub} = \varepsilon_{super} = 2.1$, which is exact the relative permittivity of silica in infrared region. The structure of 1D PhCs is shown in Fig.(\ref{Fig(1)}b), where the PhCs is consist of five layers in the unit cell with relative permittivity as $\varepsilon_A = 2.1$, $\varepsilon_B = 3.24$, $\varepsilon_C = 1.44$, $\varepsilon_B = 3.24$, $\varepsilon_A = 2.1$ and the widths of five layers are set as $d_A/2$, $d_B/2$, $d_C$, $d_B/2$, $d_A/2$, respectively. Obviously, the PhCs exhibits inversion symmetry at the central point of layer-C. This structure is called "ABCBA-model". In this work, the relative permeability of all dielectric materials is supposed to be $\mu_r = 1$. In Fig.(\ref{Fig(1)}c), in the deep sub-wavelength region with $\lambda >> a$ (where $a = d_A + d_B+ d_C$ is the length of unit cell and is set as $a = 1\mu m$), the ABCBA-model can be modelled by the same effective uniaxial medium with a relative permittivity of $\bar{\bar{\varepsilon}} = [\varepsilon_\parallel, \varepsilon_\parallel, \varepsilon_\perp]$. According to the EMT \cite{choy2015effective, yuan2023breakdown}, we have $\varepsilon_\parallel = \frac{\varepsilon_Ad_A+\varepsilon_Bd_B+\varepsilon_Cd_C}{a}$ and $\varepsilon_\perp = \frac{a}{d_A/\varepsilon_A+d_B/\varepsilon_B+d_C/\varepsilon_C}$. Transfer matrix method (TMM) is a strict and useful method when considering multi-layer structure, so that we will use TMM to calculate the properties of our model below and after.

\begin{figure*}[htb]
    \centering
    \includegraphics[width=1\linewidth]{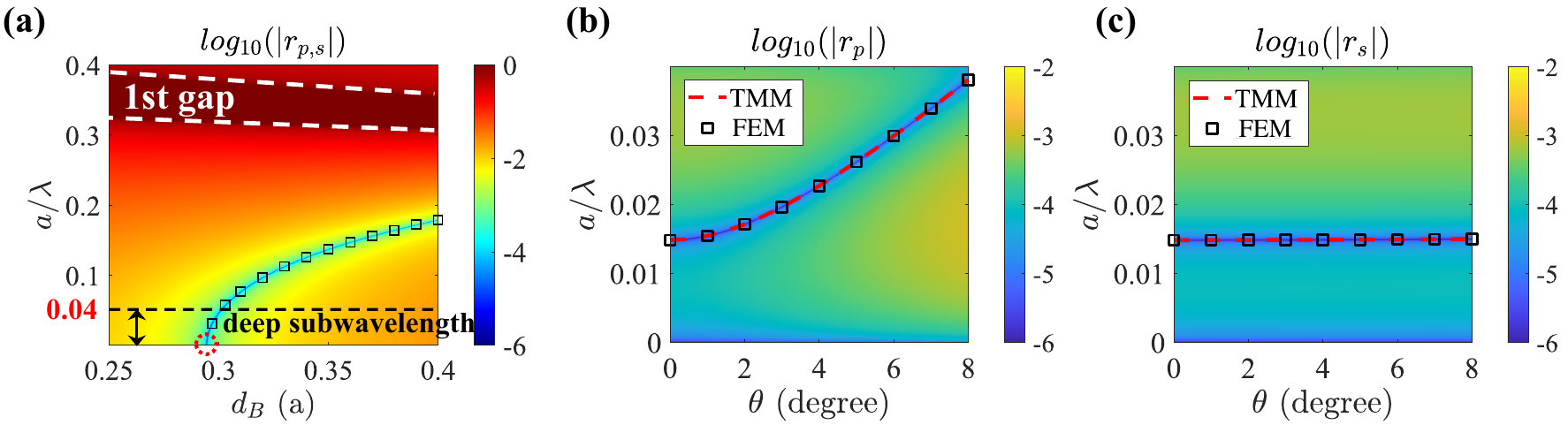}
    \caption{(a) The reflection coefficient of s- (p-) polarization when normally incident. The deep red region is the first gap and the deep blue line is the trajectory of topological singularity, which is also verified by the black squares calculated by FDTD. The topological singularity reaches zero frequency when $d_{B0} = 0.2944a$, marked by ret dots.Deep sub-wavelength is defined as $\lambda \geq 25a$, as the black dashed line in (a). (b) and (c) are the reflection coefficient of p- and s-polarization in $\{\theta, \lambda \}$ space with $d_B = 0.295a$. The pink dashed lines and black squares are the trajectory of topological singularity for both polarizations calculated by TMM and FDTD, respectively.}
    \label{Fig(2)}
\end{figure*}


First, we discuss the evolution of topological singularity in deep-subwavelength region. In this work, the deep-subwavelength is defined as $\lambda \geq 25a$. We also set $\lambda > n_{eff}Na/2$, where $n_{eff}$ is the effective index of our model, so that there is no Bragger reflections resonant mode \cite{kalozoumis2018finite} in this region. In Fig.(\ref{Fig(2)}a), we depict the reflection coefficient (in log) in $\{d_B, \lambda\}$ space when normally incident, where $d_A = 0.2a$, $d_C = a-d_A-d_B$ and $N = 8$. The deep red region indicates the first gap, marked by white dashed lines. The deep blue lines, indicating perfect transmission and zero reflection, is the trajectory of topological singularity in $\{d_B, \lambda\}$ space. To verify the correctness of our results from TMM, we use the finite element method (FEM) as the numerical experiments to obtain the trajectory of topological singularity (marked by the black squares), which agrees well with our results from TMM. We find that, with the decrease of $d_B$, the singularity will move towards lower frequency (higher wavelength) region and it reaches zero frequency (infinite wavelength) at $d_{B0} = 0.2944a$, which satisfies \cite{liu2023evolution}:

\begin{equation}\label{eq(2)}
  \varepsilon_{super} = \frac{\varepsilon_Ad_A+\varepsilon_Bd_B+\varepsilon_Cd_C}{a}
\end{equation}

Further decreasing $d_B$, the topological singularity will evolve into the pure imaginary frequency region, details are shown in the supplementary section S1. When the light is obliquely incident with angle $\theta_i$, the reflection properties of p-polarization and s-polarization are different. In Fig.(\ref{Fig(2)}b) and Fig.(\ref{Fig(2)}c), we depict the reflection coefficient of p-polarization and s-polarization in deep sub-wavelength region ($a/\lambda \leq 0.04$) with $d_B = 0.295a$. The red dashed lines and black squares are the trajectory of topological singularity for both polarizations calculated by TMM and FEM, which agree well with each other. we show that the singularity trajectories of s-polarization are more flat than those of p-polarization because the effective dielectric constant $\varepsilon_\parallel =2.1$ is larger than $\varepsilon_\perp = 1.44$  \cite{liunearly}. Therefore, the singularities for p-polarization and s-polarization will gradually separate as the incident angle increases. The zero-reflection of topological singularities in deep-subwavelength region provides a perfect platform to research photonic spin-Hall effect.

Next, we will investigate the PSHS of topological singularity in deep sub-wavelength region and compare with the effective medium model. According to the three-dimensional rotation matrix, near the topological singularity with $r_{s,p} \rightarrow 0$, the approximation of Eq.\eqref{eq(1)} fails, and the transverse spin-shift $\Delta y_{s,p}^{\pm}$ of reflected Gaussian beam with finite waist radius $w_0$ can be approximately rewritten as \cite{wang2013spin}:
\begin{equation}\label{eq(3)}
  \Delta y^{\pm}_{s,p} = \pm \frac{\delta_{s,p}}{1+\frac{|\Delta_{s,p}|^2+|\delta_{s,p}|^2}{w_0^2}}
\end{equation}
where,
\begin{equation}\label{eq(4)}
  \Delta_{s,p} = -\frac{2\pi}{\lambda}\frac{\partial ln(r_{p,s})}{\partial\theta}
\end{equation}
\begin{equation}\label{eq(5)}
  \delta_{s,p} = -\frac{2\pi}{\lambda}(1 + \frac{r_{p,s}}{r_{s,p}})
\end{equation}
Detailed derivations are shown in the supplementary section S2. According to Eq.\eqref{eq(3)}, it is obvious that the PSHS of p-polarization and s-polarization are determined by the relative valus of $|\Delta_{p,s}|$ and $|\delta_{p,s}|$.

\begin{figure*}[htb]
    \centering
    \includegraphics[width=0.9\linewidth]{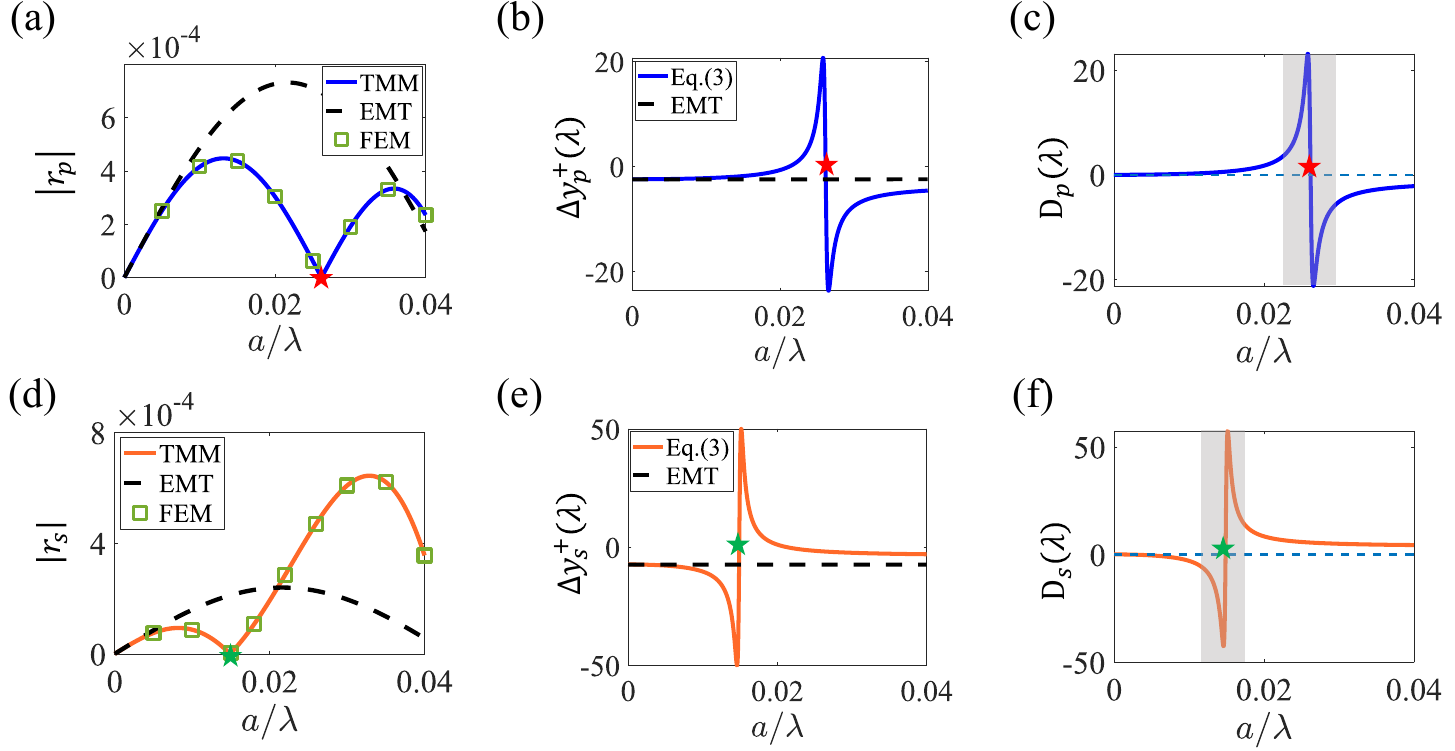}
    \caption{The spin-Hall shift for both-polarizations and the breakdown of effective medium theory in deep-subwavelength region. (a)-(c) for p-polarization and (d)-(f) for s-polarization. (a) and (d) are the reflection coefficient $|r|$ calculated by TMM (blue solid lines), EMT (black dashed lines) and FEM (green squares) for p-polarization and s-polarization. (b) and (e) are the spin Hall shift $\Delta y_{s,p}^+$ of reflected right-handed CP Gaussian beam $|+\rangle$ (units of $\lambda$) calculated by Eq.\eqref{eq(3)} and EMT for p-polarization and s-polarization. (c) and (f) are the difference of spin Hall shift between Eq.\eqref{eq(3)} and EMT for p-polarization and s-polarization. The position of topological singularity is marked by red and green pentagram for p-polarization and s-polarization, respectively.}
    \label{Fig(3)}
\end{figure*}

In Fig.(\ref{Fig(3)}a) and Fig.(\ref{Fig(3)}d), the reflection coefficient $|r|$ with incident angle $\theta_i = 5^\circ$ for p-polarization and s-polarization are calculated by TMM (solid lines) and verified by FEM (squares), respectively. The results from TMM agree well with FEM. We show that there is a topological singularity at $a/\lambda = 0.26$ for p-polarization (marked by red pentagram) and a topological singularity at $a/\lambda = 0.15$ for s-polarization (marked by green pentagram). Besides, the reflection coefficients $|r_{p,s}|$ calculated by EMT are also calculated in Fig.(\ref{Fig(3)}a) and Fig.(\ref{Fig(3)}d), we show that there is no zero-reflection for EMT, so that the breakdown of EMT is occured in the deep-subwavelength region. We further calculate the PSHS of reflected right-handed CP Gaussian beam $|+\rangle$ for both polarizations in Fig.(\ref{Fig(3)}b) and Fig.(\ref{Fig(3)}e). The solid lines are calculated by Eq.\eqref{eq(3)} and dashed lines are calculated by EMT. We show that, near singularity, huge PSHS is achieved for both-polarizations, which reaches to $24\lambda$ for p-polarization and $50\lambda$ for s-polarization with waist radius $w_0 = 100\lambda$. Besides, the PSHS for both polarization is near zero at the singularity due to the fact that $\delta_{s,p} >> w_0$ and $\delta_{s,p} >> w_0$ at singularity. While for EMT, it remains tiny due to the absence of singularity, which is accordance with Eq.\eqref{eq(1)}. In Fig.(\ref{Fig(3)}c) and Fig.(\ref{Fig(3)}f), we further calculate the difference of PSHS calculated between Eq.\eqref{eq(3)} and EMT (i.e., $D_{s,p} = \Delta_{y,s/p}^{Eq(3)}-\Delta_{y,s/p}^{EMT}$) for p-polarization and s-polarization, respectively. We show a dramatic breakdown of EMT approximation is occurred near the topological singularity in the deep-subwavelength, as the gery shadows shown in the pictures. Away from the topological singularity, the EMT still holds in deep-subwavelength region.

In this section, we show that, due to the existing of topological singularity in deep-subwavelength, huge PSHS for dual-polarizations can be achieved and the EMT is breakdown in this case. We note that when the topological singularity is away from sub-wavelength region (such as, moving into the higher frequency region or moving into the pure imaginary frequency domain), the EMT will be applicable again. The details can be seen in the supplementary section S1. 

\section{\label{sec:3} Deep-subwavelength detection based on the topological singularity}
In the previous section, we have shown that the PSHS reaches zero at singularity and huge PSHS for both polarizations can be achieved near the singularity. In this section, we will show that the topological singularity is sensitive to the structure and material parameters and dramatically changes of PSHS near the topological singularity will be beneficial to design dual-polarizations spin-sensitive thickness and refractive index sensing devices. 

\begin{figure*}[htb]
    \centering
    \includegraphics[width=1\linewidth]{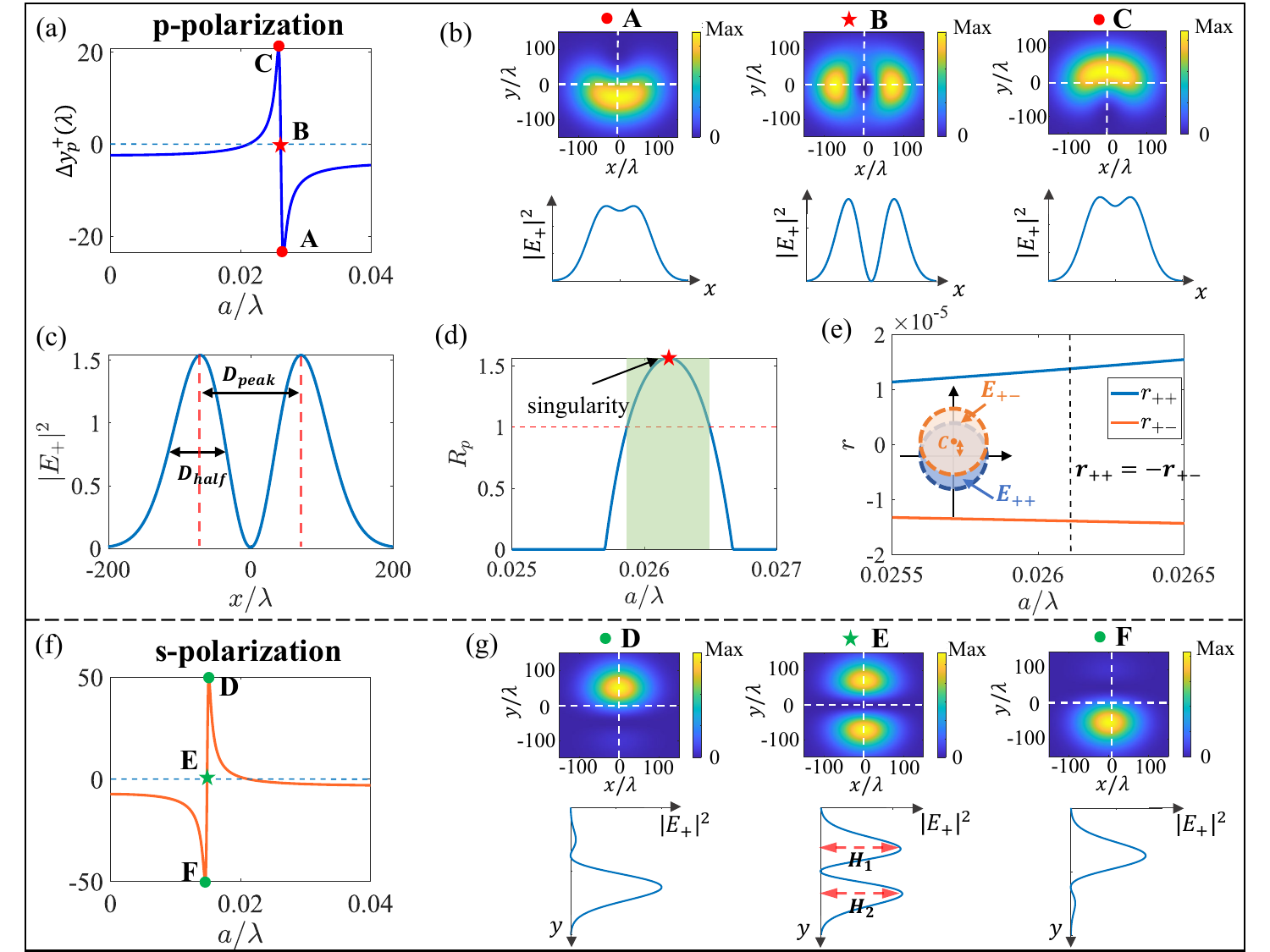}
    \caption{(a) Photonic spin Hall shift for p-polarization. (b) The 2D and 1D filed distribution of the reflected right-handed CP $|+\rangle$, marked by A, B and C in (a). The waist radius of Gaussian beam is set as $w_0 = 100\lambda$. (c) The normalized filed distribution of B-point in (b) among x-direction with $y = 0$. (d) The resolution ratio $R_p$ for p-polarization. Green shadow indicates $R_p > 1$. (e) Fresnel coefficients $r_{++}$ and $r_{+-}$ for the CP plane waves. $r_{++} = -r_{+-}$ at the topological singularity. Inset: the illustration of interfere between abnormal and normal light.(f) Photonic spin Hall shift for s-polarization. (b) The 2D and 1D filed distribution of the reflected $|+\rangle$ CP field, marked by D, E and F in (f).}
    \label{Fig(4)}
\end{figure*}

For p-polarization, the PSHS is shown in Fig.(\ref{Fig(4)}a). The minimum values, zero and maximum values of PSHS are marked by A, B and C, where B is the position of topological singularity with $\lambda = 38.2a$. The filed distribution of reflected CP Gaussian beam can be written as:

\begin{equation}\label{eq(6)}
\begin{split}
  &|E(x,y)|^2 = \exp(-2\frac{x^2+y^2}{w_0})\times \frac{|r_{s,p}|^2}{2} \{[(\frac{2\Delta_{s,p}}{w_0})^2x^2 + \\&
  (\frac{2\delta_{s,p}}{w_0})^2y^2]|+\rangle + [(\frac{2\Delta_{s,p}}{w_0})^2x^2 - (\frac{2\delta_{s,p}}{w_0^2})^2y^2]|-\rangle\}
\end{split}
\end{equation}
Detailed derivations are shown in the supplementary section S3. According to Eq.\eqref{eq(6)}, near the topological singularity, the reflected field distributions of p-polarization and s-polarization are determined by relative values between $|\Delta_{s,p}|$ and $|\delta_{s,p}|$ (also see in the supplementary section S3). The two-dimensional (2D) and 1D field distribution of the right-handed CP Gaussian beam $|+\rangle$ marked by A, B and C are shown in Fig.(\ref{Fig(4)}b). For A (C), the field is arc-like distribution with negative (positive) transverse spin Hall shift. Interesting, for B, the field is split as dipole-like distribution in $x$ direction with near-zero transverse spin Hall shift due to the fact that $|\Delta_{s,p}| > |\delta_{s,p}|$. This phenomenon is also discovered for the Brewster angle in the single interface system \cite{ling2021revisiting} and perfect spin-conversion points in multi-layered system \cite{liunearly}. Quantitatively, in Fig.(\ref{Fig(4)}c), we depict the normalized filed distribution of B-point among x-direction with $y = 0$. There are two peaks at $x = w_0/\sqrt2$ and zero value at $x = 0$, which is consistent with Fig.(\ref{Fig(4)}b). We define that $D_{peak}$ is the distance between two peaks and $D_{half}$ is the half of one of peak. To better distinguish the case of B and A/C, we further define the ratio between $D_{peak}$ and $D_{half}$ as:
\begin{equation}\label{eq(7)}
  R_p = \frac{D_{peak}}{D_{half}}
\end{equation}
where, $R_p$ is the resolution factor for p-polarization. When $R_p > 1$, the field can be regarded as dipole-like distribution. When $R_p < 1$, the field can be regarded as arc-like distribution. We mention that although reflected CP Gaussian beam is rather weak, it can be measured by signal enhancement technique experimentally \cite{hosten2008observation, qin2009measurement, dai2020direct}. In Fig.(\ref{Fig(4)}d), the ratio $R_p$ is plotted in the deep sun-wavelength region, where green shadow indicates $R_p > 1$. We show that a narrow frequency-region near the topological singularity with  $R_p > 1$ can be regarded as dipole-like distribution.

\begin{figure*}[htb]
    \centering
    \includegraphics[width=1\linewidth]{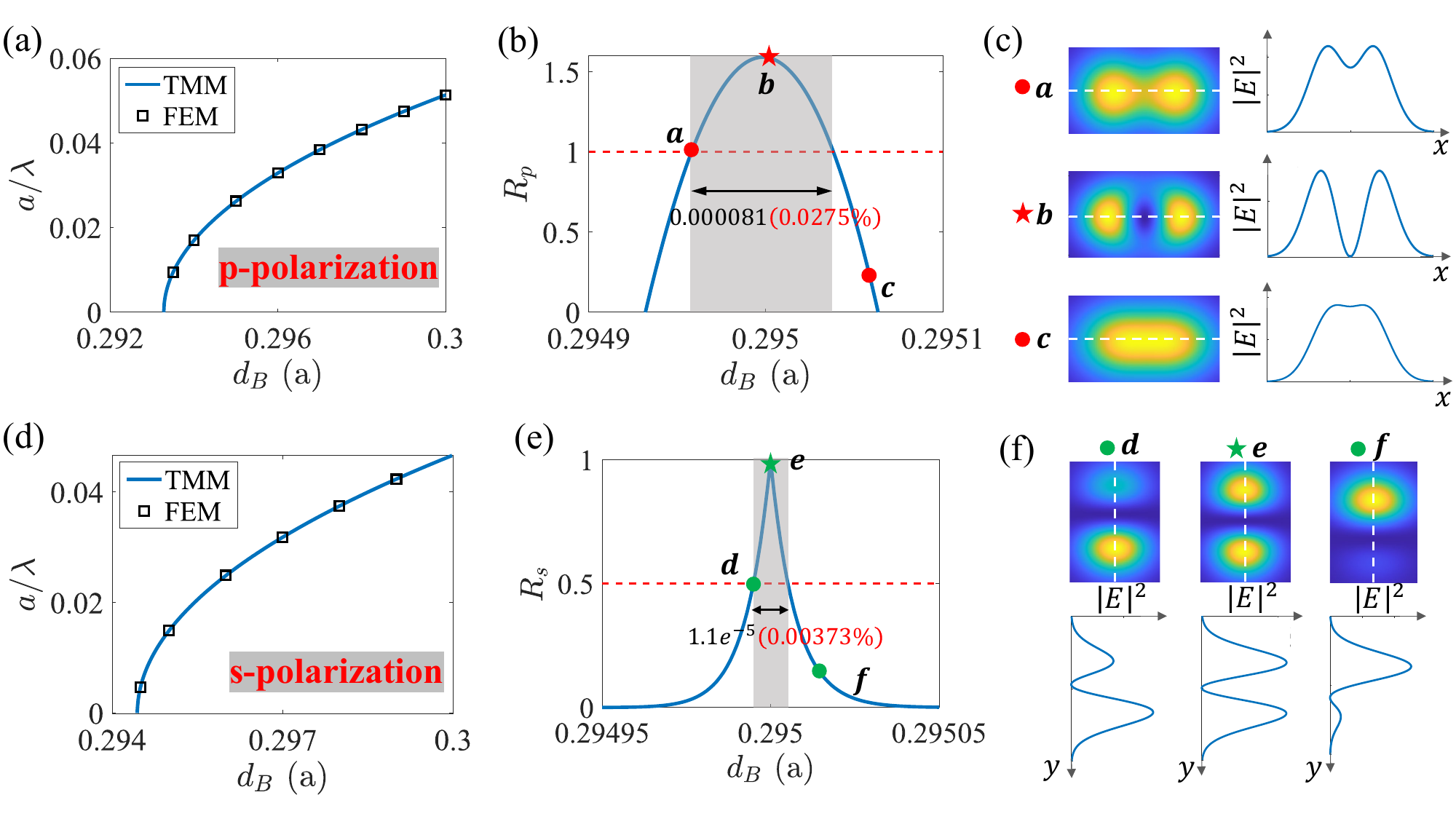}
    \caption{Thickness detection for p-polarization ((a)-(c)) and s-polarization ((d)-(f)). (a) and (d) are the trajectory of topological singularity with different $d_B$ calculated by TMM (blue lines) and FEM (black squares). (b) and (e) are resolution ratio $R_p$ (with incident wavelength $\lambda = 38.2a$) and $R_s$ (with incident wavelength $\lambda = 67.2a$) with different $d_B$, where the grey shadow indicates $R_p \geq 1$ and $R_s \geq 0.5$. (c) and (f) are the 1D and 2D filed distribution of the reflected right-handed CP Gaussian beam $|+\rangle$, marked by a-f in (b) and (e).}
    \label{Fig(5)}
\end{figure*}

The similar phenomenon is also shown for s-polarization. In Fig.(\ref{Fig(4)}f), we also depict the PSHS for s-polarization in deep-sub-wavelength region, where the maximum values, zero and minimum values of PSHS are marked by D, E and F, in which E is the position of topological singularity with $\lambda = 67.2a$. The 2D and 1D field distribution of the reflected right-handed CP Gaussian beam $|+\rangle$ marked by D, E and F are shown in Fig.(\ref{Fig(4)}g). For D (F), the field is circular-like distribution with positive (negative) transverse spin Hall shift. Interesting, we show that a split of reflected field at singularity (E) is occurred among y-direction, where two peaks $H_1$ and $H_2$ are located at $y = \pm w_0/\sqrt2$. This phenomenon is due to the fact that $|\Delta_{s,p}| < |\delta_{s,p}|$ in this case (details are shown in the supplementary materials). The quantitatively distinguish the reflected field of E and D/F, we further define the resolution factor for s-polarization as:
\begin{equation}\label{eq(8)}
  R_s = \frac{min(H_1,H_2)}{max(H_1,H_2)}
\end{equation}
When $R_s > 0.5$, the field can be regarded as dipole-like distribution. When $R_s < 0.5$, the field can be regarded as circular-like distribution.

Next, we will give the illustration of the split field distribution at topological singularity physically. The p-polarization can be written in the CP basis:
\begin{equation}\label{eq(9)}
  |p\rangle = \frac{1}{2}(|+\rangle+|-\rangle) 
\end{equation}
Under paraxial approximation, the reflected CP Gaussian beam can be obtained as:
\begin{equation}\label{eq(10)}
\begin{split}
  &|+\rangle_r = E^r_{++}+E^r_{+-} \approx r_{++} |+\rangle_i + r_{+-} |-\rangle_i \\&
  |-\rangle_r = E^r_{--}+E^r_{-+}  \approx r_{--} |-\rangle_i + r_{-+} |+\rangle_i
\end{split}
\end{equation}
where $r_{++} = \frac{r_p+r_s}{2}$ and $r_{+-} = \frac{r_p-r_s}{2}$ are the Fresnel coefficients for the CP plane waves. $E^r_{++}$ ($E^r_{--}$) is the reflected spin-maintained normal light with zero spin Hall shift and $E^r_{+-}$ ($E^r_{-+}$) is reflected the spin-flipped abnormal light with spin Hall shift $\Delta y_\pm = \mp \frac{2\cot(\theta^i)}{k_0}\approx5\lambda$ in our case. The spin Hall shift is rather tiny comparing with the waist $w_0 = 100\lambda$. At the singularity for p-polarization with $r_p = 0$, we obtain that $r_{+-} = -r_{++}$. Therefore, according to Eq.\eqref{eq(10)}, the reflected normal light $E_{++}$ and abnormal light $E_{+-}$ are destructive interference near the origin, as the inset shown in Fig.(\ref{Fig(4)}e). The destructive interference between spin-maintained normal light and spin-flipped abnormal light results the split field distribution.

In above, we show that near the topological singularity, the field will undergo a dramatic variation from arc-like to dipole-like to arc-like mathematically and physically. Practically, the position of the singularity are closely related to the structure properties (i.e., thickness) and material properties (i.e., dielectric constant), so it has broad prospects in the field of precious detections.

The first device is the sensitive sub-nanometer thickness detection. In Fig.(\ref{Fig(5)}a) and Fig.(\ref{Fig(5)}d), the trajectory of topological singularity for p-polarization and s-polarization with the thickness of layer-B are calculated by TMM (solid lines) and verified by FEM (squares), which agrees well with each other. When $d_B$ is increased from $0.295a$ to $0.296a$, the singularity will be from $\lambda = 38.2a$ to $\lambda = 30.4a$ for p-polarization and from $\lambda = 67.2a$ to $\lambda = 40.2a$ for s-polarization. The sensitivity of the thickness sensor can be described by $S = \Delta\lambda/\Delta d$, where $\Delta\lambda$ is the wavelength shift value of the singularity caused by the thickness change $\Delta d$. The $S$ values of p-polarization and s-polarization reach $7800$ RIU$^{-1}$ and $27000$ RIU$^{-1}$, respectively. Therefore, the topological singularity is very sensitive to the $d_B$ for both-polarizations. Additionally, in Fig.(\ref{Fig(5)}b) and Fig.(\ref{Fig(5)}e), we further depict the resolution ratio $R_p$ and $R_s$ with the change of $d_B$, in which the grey shadow indicates $R_p \geq 1$ and $R_s \geq 1$. The 2D and 1D field distributions are shown in Fig.(\ref{Fig(5)}c) and Fig.(\ref{Fig(5)}f). For p-polarization, we show that, in the region with $R_p > 1$, the field is well defined as spilt dipole-like distribution (shown in Fig.(\ref{Fig(5)}c)), which can be easily and clearly distinguished from the case $c$ in Fig.(\ref{Fig(5)}b). So a sensitive thickness detection can be achieved by the reflection field shape of Gaussian beam. Therefore, in Fig.(\ref{Fig(5)}b), we show that the well defined dipole-like distribution region with $R_p>1$ is very sensitive to the thickness of $d_B$, where we can distinguish $0.0275\%$ change of $d_B$ by the reflected field distribution. For the structure with the lattice constant $a = 1\mu m$ that we set before, we can detect $0.081 nm$ change of the thickness of layer-B, which can be applied into sub-nanometer thickness error sensor. The sensitive thickness detection can be also applied for s-polarization. As shown in Fig.(\ref{Fig(5)}e), we can even distinguish $0.00373\%$ ($\sim0.011 nm$) change of $d_B$ by the shape of reflected field distribution.

\begin{figure*}[htb]
    \centering
    \includegraphics[width=1\linewidth]{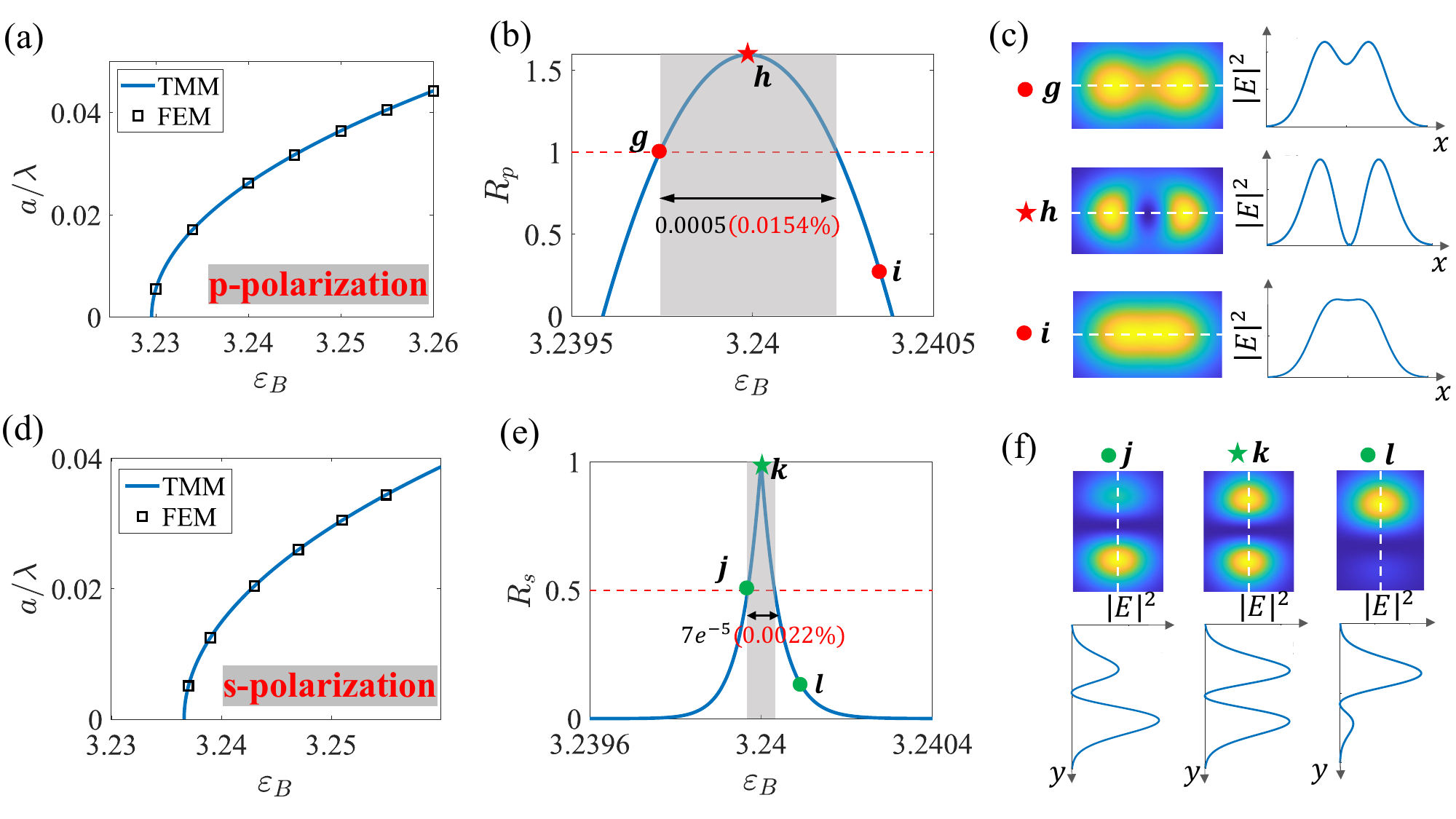}
    \caption{Dielectric constant detection for p-polarization ((a)-(c)) and s-polarization ((d)-(f)). (a) and (d) are the trajectory of topological singularity with different $\varepsilon_B$ calculated by TMM (blue lines) and FEM (black squares). (b) and (e) are resolution ratio $R_p$ (with incident wavelength $\lambda = 38.2a$) and $R_s$ (with incident wavelength $\lambda = 67.2a$) with different $\varepsilon_B$, where the grey shadow indicates $R_p \geq 1$ and $R_s \geq 0.5$. (c) and (f) are the 1D and 2D filed distribution of the reflected right-handed CP Gaussian beam $|+\rangle$, marked by g-l in (b) and (e).}
    \label{Fig(6)}
\end{figure*}

The second device is the sensitive dielectric constant $\varepsilon$ sensor. In Fig.(\ref{Fig(6)}a) and Fig.(\ref{Fig(6)}d), the trajectory of topological singularity for p-polarization and s-polarization with the dielectric constant of layer-B ($\varepsilon_B$) are calculated by TMM (solid lines) and verified by FEM (squares). We show that the topological singularity is also sensitive to $\varepsilon_B$. 
When $\varepsilon_B$ is increased from $3.24$ to $3.241$, the singularity will be from $\lambda = 38.2a$ to $\lambda = 35.6a$ for p-polarization and from $\lambda = 67.2a$ to $\lambda = 58.9a$ for s-polarization. The sensitivity of the dielectric constant sensor can be described by $S = \Delta\lambda/\Delta \varepsilon$, where $\Delta\lambda$ is the wavelength shift value of the singularity caused by the dielectric constant change $\Delta \varepsilon$. The $S$ values of p-polarization and s-polarization reach $2600\mu m/$ RIU$^{-1}$ and $8300\mu m/$ RIU$^{-1}$, respectively. Further more, in Fig.(\ref{Fig(6)}b) and Fig.(\ref{Fig(6)}e), we further depict the resolution ratio $R_p$ and $R_s$ with the change of $\varepsilon_B$, in which the grey shadow indicates $R_p \geq 1$ and $R_s \geq 1$. The 2D and 1D field distributions are shown in Fig.(\ref{Fig(6)}c) and Fig.(\ref{Fig(6)}f). We shown that a sensitive dielectric constant sensor can be achieved by the reflection field shape of Gaussian beam. In Fig.(\ref{Fig(6)}b) for p-polarization, we can distinguish $0.0154\%$ ($\sim 0.0005$) change of $\varepsilon_B$ by the reflected field distribution. In Fig.(\ref{Fig(6)}e) for s-polarization, we can even distinguish $0.0022\%$ ($\sim 0.00007$) change of $\varepsilon_B$ by the reflected field distribution. We mention that the sensitive dielectric constant sensor can also be applied to the temperature detection and power detection if the material is the nonlinear material. We also note that the topological singularity is sensitive to the dielectric constant of the substrate and superstrate, which can further applied into medical diagnosis\cite{tseng2020dielectric}, environmental detection\cite{tittl2018imaging}, etc. Details are shown in the supplementary section S4.

\begin{figure*}[htb]
    \centering
    \includegraphics[width=0.9\linewidth]{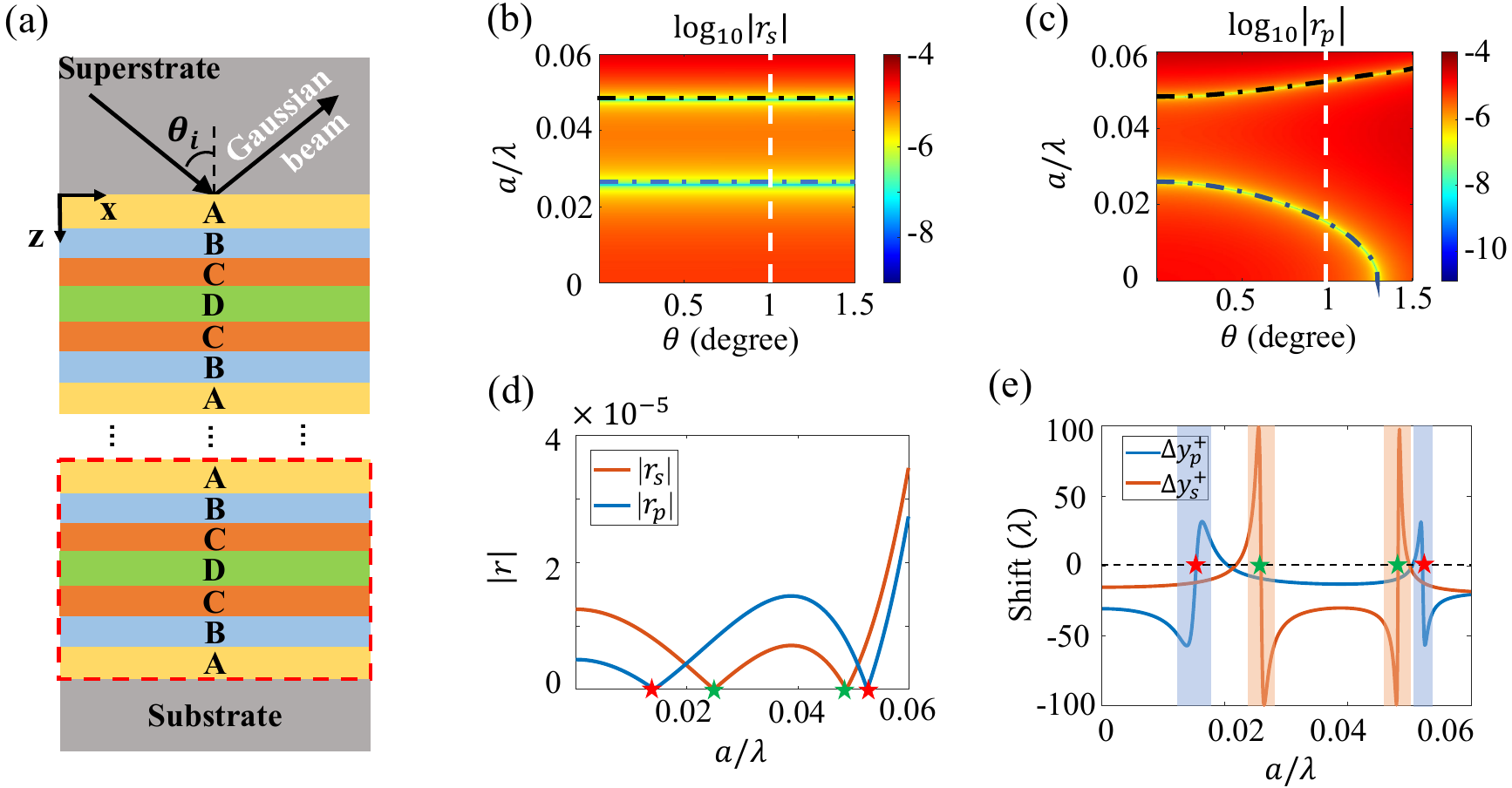}
    \caption{(a) The multi-layers ABCDCBA-model. (b) and (c) are the log of reflection coefficient in $\{\theta, \lambda\}$ space for s-polarization and p-polarization. The trajectories of topological singularities are marked by solid-dashed lines. (d) The reflection coefficient of finite PhCs for p-polarization (blue line) and s-polarization (orange line) with cell number $N = 2$ and incident angle $\theta = 1^\circ$. Topological singularities are marked by red pentagrams for p-polarization and blue pentagrams for s-polarization. (e) Photonic spin Hall shift for p-polarization (blue line) and s-polarization (red line). The waist radius is set as $w_0 = 200\lambda$. There are two detection channels for both polarizations, marked by blue and orange shadows.}
    \label{Fig(7)}
\end{figure*}

In this section, we illustrate that the reflected CP filed is split at singularity numerically and physically. Meanwhile, the topological singularity is rather sensitive to the thickness and dielectric constant in the deep-subwavelength, which can be applied to the single-channel sensitive detection of the thickness and dielectric constant for both-polarizations. Practically, this design can be applied into the manufacturing error detection, temperature detection and nonlinear power detection.

\section{\label{sec:4} Dual-polarizations multi-channels detection in subwavelength region}

\begin{figure*}[htb]
    \centering
    \includegraphics[width=1\linewidth]{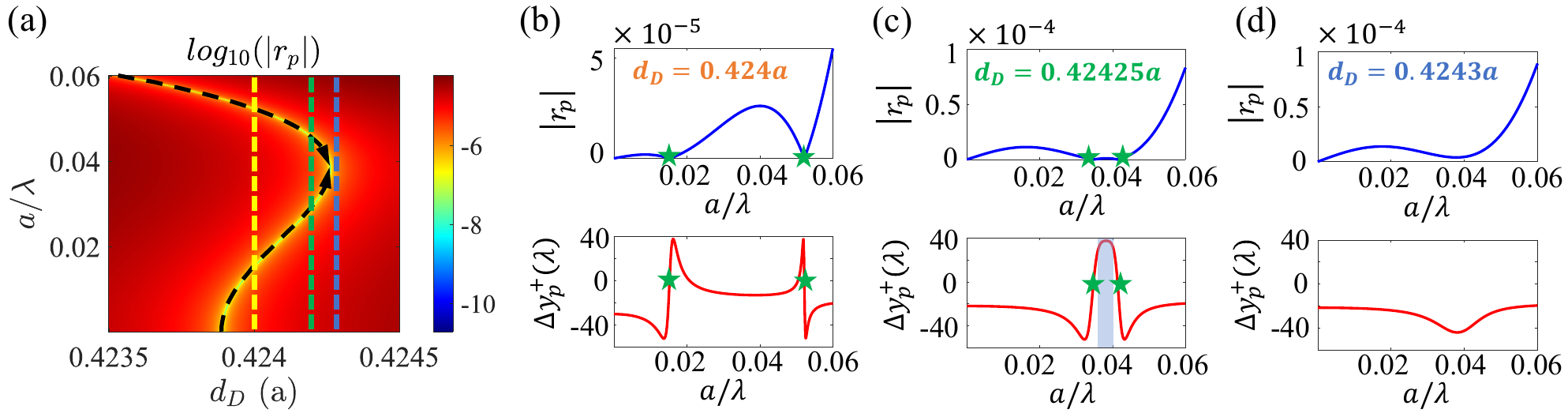}
    \caption{(a) The log of reflection coefficient in $\{d_D, \lambda\}$ space for p-polarization with $\theta = 1^\circ$. The trajectories of topological singularities are marked by black solid-dashed lines. Two singularities will annihilate at $d_D = 0.43428 a$. (b)-(d) The reflection coefficient (upper) and PSHS (downer) for p-polarization with $d_D = 0.424a$, $d_D = 0.42425a$ and $d_D = 0.4243a$, respectively. Blue shadow in (c) indicates huge-PSHS platform for p-polarization.}
    \label{Fig(8)}
\end{figure*}

In the above sections, we have shown that for "ABCBA-model", there is a singularity in deep-subwavelength for both-polarizations so that single-channel dual-polarizations sensitive detection can be designed utilizing the topological singularity. In this section, we will show that for more complicated layered structure, more topological singularities can be achieved in the subwavelength region for both polarizations so that multi-channels dual-polarizations detection can be designed.

The more complicated layered structure is shown in Fig.(\ref{Fig(7)}a), where the PhCs is consist of seven layers in the unit cell with relative permittivity as $\varepsilon_A = 5.999$, $\varepsilon_B = 4$, $\varepsilon_C = 9$, $\varepsilon_D = 4.8$, $\varepsilon_C = 9$, $\varepsilon_B = 4$, $\varepsilon_A = 5.999$ and the thickness of seven layers are set as $d_A/2 = 0.041a$, $d_B/2 = 0.1a$, $d_C/2 = 0.15a$, $d_D = 0.418$, $d_C/2 = 0.15a$, $d_B/2=0.1a$, $d_A/2=0.041a$, where $a = d_A+d_B+d_C+d_D$ is the length of unit cell. The structure is called "ABCDCBA-model" \cite{liu2023evolution}. In Fig.(\ref{Fig(7)}b) and Fig.(\ref{Fig(7)}c), the log of reflection coefficients are depicted in $\{\theta, \lambda\}$ space for s-polarization and p-polarization. We find that there are two singularities in the subwavelength region with $\lambda\geq15a$ (i.e., $a/\lambda <= 0.06$). The trajectories of topological singularities are labeled by blue and black dashed-solid lines. For s-polarization in Fig.(\ref{Fig(7)}b), the trajectories of two singularities are nearly fixed with the increase of $\theta$. While for p-polarization in Fig.(\ref{Fig(7)}c), two singularities will gradually separate with each other with the increase of $\theta$. In Fig.(\ref{Fig(6)}d), the reflection coefficient of finite PhCs with cell number $N = 2$ and incident angle $\theta = 1^\circ$ for p-polarization (blue line) and s-polarization (orange line) are calculated. The positions of topological singularities are marked by red pentagrams (i.e., $\lambda = 71a$ and $\lambda = 19a$) for p-polarization and green pentagrams (i.e., $\lambda = 40a$ and $\lambda = 21a$) for s-polarization, respectively. In Fig.(\ref{Fig(6)}e), we further calculate the PSHS for p-polarization (blue line) and s-polarization (red line) in subwavelength region. We show that, two channels dual-polairzations huge PSHS can be achieved for "ABCDCBA-model", in which PSHS reaches $50a$ for p-polarization and $100a$ for s-polarization with waist radius $w_0=200a$. Meanwhile, near the four topological singularities (two singularities for s-polarization and two singularities for p-polarization), the PSHS changes dramatically, which means that there are two sub-wavelength detection channels for s-polarization (marked by orange shadows) and two subwavelength detection channels p-polarization (marked by blue shadows), respectively. Additionally, more singularities can be achieved in more complicated layered structure or periodic modulation model \cite{liu2023evolution} so that more detection channels can be designed applied to multi-frequency multiplexing applications. These dual-polarizations multi-channels can be also applied into the thickness, dielectric constant and background sensor, ect. 

In the above devices, huge PSHS can be achieved near the topological singularity, which is very sensitive to the incident frequency (wavelength) and less robust. Here, except for these applications, a broadband platform with huge-PSHS can be realized for both-polarizations in subwavelength region in the "ABCDCBA-model". In Fig.(\ref{Fig(8)}a), the log of reflection coefficient with $\theta_i=1^\circ$ are depicted in $\{d_D, \lambda\}$ space for p-polarization, in which the trajectories of singularities are marked by black dashed lines. We show that with the increasing of $d_D$, two singularities will gradually approach each other and they will annihilate each other in real frequency domain at $d_D = 0.45428a$ and then evolve into the complex frequency domain \cite{liu2023evolution}. In Fig.(\ref{Fig(8)}b)-Fig.(\ref{Fig(8)}d), The reflection coefficients (upper) and PSHSs (downer) for p-polarization are depicted with $d_D = 0.424a$, $d_D = 0.42425a$ and $d_D = 0.4243a$, respectively, labeled by dashed lines in Fig.(\ref{Fig(8)}a). The topological singularities are marked by green pentagrams. The upper panels in Fig.(\ref{Fig(8)}b)-Fig.(\ref{Fig(8)}d) show the singularities approach-annihilation process with the increase of $d_D$. In the downer panels in Fig.(\ref{Fig(8)}b)-Fig.(\ref{Fig(8)}d), we shown that huge PSHS ($\sim 50\lambda$) can be achieved near two topological singularities. More interestingly, we find that when singularities approach each other, the PSHS between two singularities will become flatter. When $d_D = 0.42425a$, there is a wide wavelength-range $\lambda = 25a\sim28a$ platform with huge-PSHS reaches to $\Delta_p^+ \geq 37\lambda$, as the blue shadow shown in the downer panel of Fig.(\ref{Fig(8)}c), which provides a perfect platform to research PSHS and apply into the detection devices. Further increasing $d_D$, two singularities will annihilate each other and the huge-PSHS will disappear due to the disappearing of topological singularities in the real frequency domain. The huge PSHS platform with wide frequency range can also be realized for s-polarizations in subwavelength region, details are shown in the supplementary section S5.

\section{\label{sec:5}CONCLUSION}
In this work, we study the PSHS based on the topological singularity in deep-subwavelength in 1D multi-layer PhCs with inversion symmetry. First, huge PSHS near the topological singularity can be achieved for s-polarization ($\sim 50\lambda$) and p-polarization ($\sim 25\lambda$) with waist radius $w_0 = 100a$ in deep-subwavelength, which breaks the EMT. Second, the split reflected field distribution with zero PSHS at the topological singularity is demonstrated numerically and physically, which is resulted from the interfere of the spin-maintained normal light and spin-flipped abnormal light. Third, based on this exotic property, single-channel dual-polarizations thickness and dielectric constant sensor can be designed. Finally, by applying more complex "ABCDCBA-model", two singularities can be achieved in the subwavelength so that multi-channels detection can be designed. Further more, a huge PSHS platform with wide frequency range can be realized for both-polarizations in subwavelength region. Theoretically, this work builds the bridge between topology study of 1D PhCs and light-matter interaction in deep-subwavelength region. In practice, our model provides an unprecedented platform to research photonic spin-Hall effect, which can been applied in many applications, such as optical metrology, sensing and super-resolution imaging.

\section*{Methods}
\textbf{Numerical simulation}. Finite element method verifications were calculated using commercial software (COMSOL Multiphysics). For reflection coefficient, a unit cell with Bloch boundary conditions along the $x-$ and $y-$ axes and ports along the z-axis was simulated.

\section*{Acknowledgement}
This work is supported by National Natural Science Foundation of China (Grant No. 12174073).

\section*{Data availability}
All data that support the findings of this study are included within the article (and any supplementary files).

\section*{Conflict of interest}
The authors declare that they have no conflict of interest.

\bibliographystyle{apsrev4-2}
\bibliography{manuscript}

\end{document}